\documentstyle[12pt]{article}
\def\simgt{\stackrel{>}{{}_\sim}}
\def\be{\begin{equation}}
\def\ee{\end{equation}}
\def\bear{\be\begin{array}}
\def\eear{\end{array}\ee}
\def\bea{\begin{eqnarray}}
\def\eea{\end{eqnarray}}

\def\baselinestretch{1}
\begin{document}
\catcode`@=11
\newtoks\@stequation
\def\subequations{\refstepcounter{equation}%
\edef\@savedequation{\the\c@equation}%
  \@stequation=\expandafter{\theequation}
  \edef\@savedtheequation{\the\@stequation}
  \edef\oldtheequation{\theequation}%
  \setcounter{equation}{0}%
  \def\theequation{\oldtheequation\alph{equation}}}
\def\endsubequations{\setcounter{equation}{\@savedequation}%
  \@stequation=\expandafter{\@savedtheequation}%
  \edef\theequation{\the\@stequation}\global\@ignoretrue

\noindent}
\catcode`@=12
\begin{titlepage}
\title{{\bf The problem of the stabilization of the dilaton in string 
theories} 
\thanks{Talk given in the Fourth Conference on Supersymmetry
(SUSY 96), Univ. of Maryland at College Park,  May 29 - June 1, 1996.
Research supported in part by: the CICYT, under
contracts AEN95-0195; the European Union,
under contract CHRX-CT92-0004.}
}
\author{ {\bf J.A. Casas\thanks{On leave of absence from Instituto de
Estructura de la Materia CSIC, Serrano 123, 28006 Madrid, Spain.}
${}^{ {\footnotesize,\S}}$}\\
\hspace{3cm}\\
${}^{\footnotesize\S}$ {\small Santa Cruz Institute for Particle Physics}\\
{\small University of California, Santa Cruz, CA 95064, USA}}
\date{}
\maketitle
\def\baselinestretch{1.15}

\begin{abstract}
\noindent
The crucial problem of how the dilaton field is stabilized at a 
phenomenologically acceptable value in string theories remains
essentially unsolved. We show that the usual scenario of assuming
that the dilaton is fixed by the (SUSY breaking) dynamics of just
the dilaton itself (dilaton dominance scenario) is {\em inconsistent} 
unless the K\"ahler potential receives very important perturbative or 
non-perturbative contributions. Then, the usual predictions about
soft breaking terms are lost, but still is possible to derive
model-independent predictions for them.

\end{abstract}

\thispagestyle{empty}

\vspace{5cm}
\leftline{SCIPP-96-33}
\leftline{IEM-FT-138/96}
\leftline{Jul 1996}
\vskip-20.cm

\rightline{SCIPP-96-33}
\rightline{IEM-FT-138/96}
\rightline{hep-th/9608010}

\end{titlepage}
\newpage
\setcounter{page}{1}

\section{Introduction}
The crucial problem of how the dilaton field, $S$, whose expectation value 
gives the tree level gauge coupling constant at the string scale
\cite{Witten}
($\langle{\rm Re} S\rangle\sim g_{string}^{-2}$) is stabilized
at the experimentally ``observed'' value, $g_{string}^{-2}\simeq 2$,
remains as a very challenging one. The dilaton potential is flat at all
orders in perturbation theory as long as SUSY remains unbroken.
On the other hand SUSY itself must be broken by non-perturbative effects,
which strongly suggests that the same dynamics that breaks SUSY is also 
responsible for the stabilization of the dilaton (see ref.\cite{racetrack}). 

In this sense a very economical and model-independent scenario is the 
so-called ``Dilaton Dominance'' \cite{Kaplu93,Brignole} one. This
consists in assuming that SUSY is broken in the dilaton sector. 
In other words, only the $F_S$ auxiliary field is to take a non-vanishing VEV.
Furthermore, the dilaton dependence of the K\"ahler potential, $K$, 
is assumed to be sufficiently well approximated by the tree-level expression, 
$K=-\log(S+\bar S)$.  These assumptions are completed with the 
phenomenologically mandatory one that the superpotential $W$ is in such
an (unknown) way that the minimum of the potential, $V$, lies at an 
acceptable value for the dilaton\footnote{Here we are assuming that the
Kac-Moody level of the gauge group is $k=1$, which is the most common 
possibility; otherwise $\langle{\rm Re} S\rangle= g_{string}^{-2}/k$.}
($\langle{\rm Re} S\rangle = g_{string}^{-2}$) 
and at vanishing cosmological constant ($V=0$).
The previous assumptions lead to some 
interesting relationships among the different soft terms,
with an automatical implementation of universality, something
which is phenomenologically welcome for FCNC reasons \cite{Dim81}.

The interest of this scenario raises two questions, which directly
concern the problem of the stabilization of the dilaton: 
\begin{description}

\item[\hspace{0.4cm}{\em i)}] Is there any
form of the superpotential $W(S)$ (preferably with some theoretical
justification) able to fulfill the previous
requirements (i.e. a minimum of the scalar potential at 
$\langle {\rm Re}S\rangle\simeq 2$ and $V=0$)? 

\item[\hspace{0.4cm}{\em ii)}] How good is expected to
be the tree-level approximation for $K(S,\bar S)$?

\end{description}

Regarding the first question, ({\em i}), I will survey in sect.2
a simple analytical argument (which has been presented in ref.\cite{casas}) 
that proves that, even if the potential
has a minimum at $\langle {\rm Re}S\rangle\simeq 2$ and $V=0$, there
must exist an additional minimum (or unbounded
from below direction) in the perturbative region of $S$-values for which
$V<0$. This can be proven for an arbitrary form $W(S)$. 
Similarly, we will show that the very existence of such a (local)
minimum is forbidden unless a really huge conspiracy of different
contributions to $W(S)$ takes place.

Concerning the second question, ({\em ii}), there are, unfortunately,
indications that the stringy non-perturbative corrections to the K\"ahler
potential may be sizeable \cite{Banksdine,Shenker}. In principle, 
this may be good news since 
it could help to avoid the above-mentioned problems. The trouble here is that
very little is known about the form of these non-perturbative corrections.
The possibilities and predictions of such a scenario are explored in
sections 3 and 4.

\section{Inconsistencies of the dilaton dominated scenario}

Taking the tree-level expression for the K\"ahler potential
\be
\label{K}
K=-\log(S+\bar S)
+\hat K(T,\bar T,\phi_I,\bar \phi_I)\;\;,
\ee
where $T,\phi_I$ denote generically all the moduli and matter fields
respectively, the scalar potential in the dilaton-dominance assumption
reads
\be
\label{V}
V=\frac{1}{2\ {\rm Re}S}\left\{\left|(2\ {\rm Re}S)W_S-W\right|^2-3\left|
W\right|^2\right\}
\ee
with $W_S\equiv \partial W/\partial S$. If the scenario is realistic, 
the previous potential should  have a minimum at a realistic value
of $S$, say $S_0$, with
\be
\label{S2}
{\rm Re}S_0=\frac{1}{g^2}\simeq 2\;\;,
\ee
where, for simplicity of notation, $g$ denotes the gauge coupling constant
at the string scale ($\sim 10^{17}$ GeV). 
In addition, the vanishing of the cosmological constant, i.e.
$V(S_0)=0$, implies
\be
\label{Lambda0}
\left|(2\ {\rm Re}S_0)W_S(S_0)-W(S_0)\right|=\sqrt{3}\left|
W(S_0)\right|
\ee
and, thus
\be
\label{ineq}
\frac{2\ {\rm Re}S_0}{\sqrt{3}+1}\leq 
\left|\frac{W}{W_S}\right|_{S=S_0}\leq
\frac{2\ {\rm Re}S_0}{\sqrt{3}-1}
\ee
Performing
the following change of variables 
\be
\label{change}
z=e^{-\beta S}\;\;\;(\beta\;{\rm arbitrary})
\ee
the physical region of $S$, i.e. ${\rm Re}S >0$, 
is mapped into the circle of radius 1 in the 
$z$-plane. The ``realistic minimum'' point, $z_0=e^{-\beta S_0}$, lies
somewhere inside the circle. In the new variable, the functions $W,W_S$
are written as
\bea
\label{Omegas}
W(S)&=&\Omega(z)
\nonumber\\
W_S(S)&=&\Omega_S(z)\equiv -\beta z \Omega '(z)
\eea
and condition (\ref{ineq}) becomes
\be
\label{ineq2}
\frac{2\ {\rm Re}S_0}{\sqrt{3}+1}\leq 
\left|\frac{\Omega(z)}{\Omega_S(z)}\right|_{z=z_0}\leq
\frac{2\ {\rm Re}S_0}{\sqrt{3}-1}
\ee
with ${\rm Re}S_0=-\log|z_0|/\beta$. Let us consider now the function
\be
\label{ro}
\rho(z)\equiv\frac{\Omega(z)}{\Omega_S(z)}\;.
\ee
Let us suppose for the moment that $\rho(z)$ is an analytical function 
with no poles inside the physical region $|z|<1$. Then, the maximum
of $|\rho(z)|$ in the region $|z|\leq|z_0|$ must necessarily occur (principle 
of maximum) at some point $z_M$ belonging to the boundary, namely 
the circle ${\cal C}\equiv\{|z|=|z_0|\}$.
If we consider the larger region
enclosed by the broader circle ${\cal C'}
\equiv\{|z|=|z_0'|\}$, with $|z_0'|>|z_0|$, the new maximum
of $|\rho(z)|$ must occur now at some point, say $z=z_1$, belonging 
to the boundary ${\cal C'}$. 
{}From (\ref{ineq2}) it is clear that at $z_1$
\be
\label{ineq3}
|\rho(z_1)|=\left|\frac{\Omega(z)}{\Omega_S(z)}\right|_{z=z_1}
> \frac{2\ {\rm Re}S_0}{\sqrt{3}+1}
\ee
Taking the radius of ${\cal C'}$ so that ${\rm Re}S_1\equiv
-\log|z_1|/\beta=\frac{\sqrt{3}-1}{\sqrt{3}+1}{\rm Re}S_0$,
we can write (\ref{ineq3}) as
\be
\label{ineq4}
\left|\frac{\Omega(z)}{\Omega_S(z)}\right|_{z=z_1}
=\left|\frac{W}{W_S}\right|_{S=S_1}
> \frac{2\ {\rm Re}S_1}{\sqrt{3}-1}
\ee
Therefore, at $S=S_1$ the potential (\ref{V}) has a negative value. 
On the other hand $S_1$ still belongs to the perturbative region
\be
\label{S1}
{\rm Re}S_1\simeq 0.27 {\rm Re}S_0
\ee
which means $\alpha\simeq 0.14$. The only way-out to the previous argument
is to allow the function $\rho(z)\equiv\frac{\Omega(z)}{\Omega_S(z)}$ to
have some pole in the region enclosed by ${\cal C'}$. But then
$\left|\frac{\Omega}{\Omega_S}\right|\rightarrow\infty$ near the pole and
necessarily $\left|\frac{\Omega}{\Omega_S}\right|>
\frac{2\ {\rm Re}S}{\sqrt{3}-1}$ at some point with non-zero $\Omega$.
Hence we arrive to the same conclusion. 

\vspace{0.3cm}
\noindent
The previous argument shows that the realistic minimum assumed to take place
in the usual dilaton-dominated scenario can {\em never} correspond to a
global minimum.
This is not certainly the most 
desirable situation.

\vspace{0.3cm}
\noindent
We can go a bit further and show that under very general assumptions,
the realistic point $S=S_0$ cannot correspond to $V(S_0)=0$.

{}From symmetry and analyticity arguments  we know \cite{Banksdine}
that the non-perturbative superpotential must take the form
$W=\sum_i d_i e^{-a_i S}$, so it is reasonable to assume that at the
realistic point ($S_0\sim 2$) $W$ is dominated by one of the terms, say
$W\sim e^{-a S}$ (as it happens for instance in usual gaugino 
condensation). Then, the vanishing of $\Lambda_{cos}$ at $S=S_0$,
eq.(\ref{Lambda0}), implies
\be
\label{conditiontree1}
\left(-a-\frac{1}{2{\rm Re}S_0}\right)^2
-\frac{3}{4({\rm Re}S_0)^2}=0\;.
\ee
{}From (\ref{conditiontree1}) we obtain $a\simeq (\sqrt{3}-1)/4$, absolutely
incompatible with the requirement of a hierarchically small SUSY breaking
(note that $\langle W\rangle\sim 1$ TeV requires $a\simeq 18$).

It could happen, however, that two o more terms of the form 
$W=\sum d_i e^{-a_i S}$
cooperate at the particular region $S\simeq 2$ to produce a more realistic
scenario\footnote{This is the mechanism of the so-called racetrack models to
generate SUSY breaking (see e.g. \cite{racetrack} and references therein).
These models, however, lead naturally to moduli dominance SUSY breaking rather
than dilaton dominance.}. 
It is interesting to show that for this
to happen in a dilaton dominated scenario 
a really huge conspiracy must take place. From eq.(\ref{ineq})
we see that the condition $V(S_0)=0$ implies
\be
\label{comp1}
|W_S| \sim |W|
\ee
Since $W_S=-\sum a_i d_i e^{-a_i S}$ and
the condition of hierarchically SUSY breaking requires 
$a_i\simgt O(10)$, it is clear that a cancellation between terms with
different exponents must occur inside $W_S$ for (\ref{comp1}) to be 
fulfilled. Moreover, one has to demand that $S_0$ corresponds to
a minimum, which implies $\partial V/\partial S=0$ and the determinant
of the Hessian matrix ${\cal H}>0$. These two conditions imply
in turn \cite{casas,racetrack}
\be
\label{comp2}
(2\ {\rm Re}S)^2|W_{SS}| = 2|W|
\ee
\be
\label{comp3}
\frac{2}{\sqrt{3}} ({\rm Re}S)^2|W_{SSS}|\sim |W_S|\sim |W|\;.
\ee
These requirements are completely unnatural since the typical sizes of 
$W_{SS}$ and $W_{SSS}$ are $a_i^2 W$, $a_i^3 W$ respectively, i.e.
much larger that $W$. Therefore, two unpleasant fine-tunings must occur
at $S_0$ to become a (local) minimum of the potential (for related work
see ref.\cite{bru}).

\vspace{0.3cm}
\noindent
To summarize the results of this section, the standard (tree-level) 
dilaton-dominated scenario
can {\em never} correspond to a global minimum of the potential at 
$V=0$. Similarly, under very general assumptions it cannot correspond
to a local minimum either, unless a really big conspiracy of different
contributions to $W(S)$ takes place. 

\section{The role of the K\"ahler potential and a predictive scenario} 

The previous results, plus the fact that the K\"ahler potential is
likely to receive sizeable string non-perturbative corrections,
strongly suggest to consider a more general scenario, as
commented in the introduction. Thus, in this section we will study
how far we can go with the usual assumption
of ``dilaton--dominance'' (i.e. only $|F_S|\neq 0$), but leaving
the K\"ahler potential arbitrary. The potential reads
\be
\label{V2}
V=K_{S\bar S}\left| F_S\right|^2
-3e^K\left|W\right|^2\;,
\ee
where $F_S=e^{K/2}\left\{
\left(K^{-1}\right)^{\bar \phi S}\left(
\partial_\phi W + W K_\phi\right)^*
\right\}$
with $\phi$ running over all the chiral fields and the subindices denoting
partial derivatives.
If we also assume vanishing cosmological constant, then
\be
\label{Fs}
\left| F_S\right|(K_{S\bar S})^{1/2}=\sqrt{3}e^{K/2}\left|W\right|=
\sqrt{3}m_{3/2}\;.
\ee
The effective low-energy soft part of the Lagrangian (in terms of the
canonically normalized fields) is given by
\bea
\label{Lsoft}
-&{\cal L}_{\rm soft}&=
\frac{1}{2}M^a_{1/2}\bar {\hat \lambda_a} \hat \lambda_a +
\sum_I m_I^2|\hat \phi_I|^2 \nonumber\\ 
&+& 
\left( A_{IJL}Y_{IJL}\hat \phi_I\hat \phi_J\hat \phi_L
+ {\rm h.c.} \right) + \cdots
\eea
The values of the gaugino masses, $M^a_{1/2}$, scalar masses,
$m_I^2$, and  coefficients of the trilinear scalar terms, 
$A_{IJL}$, can be computed using general formulae 
\cite{Kaplu93,Hall} and eqs.(\ref{V2}--\ref{Fs}) (for more details
see ref.\cite{casas})
\bea
\label{relations}
m_I^2&=&m_{3/2}^2
\nonumber\\
\left|\frac{M_{1/2}}{m_{3/2}}\right|&=&\left[\frac{3g^4}
{4K_{S\bar S}}\right]^{1/2}
\nonumber\\
\left|\frac{M_{1/2}}{A}\right|&=&\left|\frac{g^2}{2K_S}\right|\;.
\eea
Here we have assumed that the Yukawa couplings appearing in the 
original superpotential, $W$, do not depend on the dilaton $S$. This is true
at tree-level (and thus at the perturbative level) and, since they are
parameters of the superpotential, they are not likely to be appreciably 
changed at the non-perturbative level \cite{Banksdine}.
The expression for the coefficient of the bilinear term, $B$, 
depends on the mechanism of generation
of the $\mu$ term, so we prefer to leave it as an independent parameter.
Finally, in order to fulfill the phenomenological requirement of 
(approximate) universality \cite{Dim81} we have demanded that
$K_{I\bar I}$ does not get $S$-dependent contributions\footnote{
This, of course, may not occur. However, it is a common assumption
of all existing string-based models (including the usual dilaton-dominance
model). It should be noted however that the universality of the scalar
masses (unless that of gaugino masses) is something imposed, rather
than obtained from the model.}. Notice also that in the tree level 
limit $K_S=\frac{1}{2{\rm Re}S},\ K_{S\bar S}=
\frac{1}{4({\rm Re}S)^2}$, and we recover from eqs.(\ref{relations}) 
the usual tree level relations \cite{Kaplu93,Brignole}. 

In eqs.(\ref{relations}) there are three unknowns 
($m_{3/2},K_S,K_{S\bar S}$)
and three soft breaking parameters ($m_I^2, M_{1/2}, A$). On the other hand,
the non-perturbative superpotential must take the form
$W=\sum_i d_i e^{-a_i S}$, so, as explained above, 
it is reasonable to assume that at the
realistic point, $S\sim 2$, $W$ is dominated by one of the terms, say
$W\sim e^{-a S} $, as it happens for instance in usual gaugino 
condensation. Thus, the condition $\Lambda_{cos}=0$, i.e. eq.(\ref{Fs}), 
gives us a further constraint, namely
\be
\label{condition}
\left|-a+K_S\right|^2-3K_{S\bar S}= 0
\ee
which relates the values of  $K_S,K_{S\bar S}$ and $a$. Notice that 
the latter is essentially fixed by the condition of a hierarchical SUSY 
breaking ($a\simeq 18$), thus eqs.(\ref{relations}) and 
eq.(\ref{condition}) give a non-trivial scenario whose phenomenology
could be investigated.

\section{Ansatzs for non--perturbative K\"ahler potentials}

It is tempting to go a bit further in our analysis and explore the
phenomenological 
capabilities of explicit (stringy) non-perturbative effects on $K$
which have been suggested in the literature\footnote{For other attempts
in this sense, see ref.\cite{nilles}.}.

{}From the arguments explained in ref.\cite{Banksdine}, it is enough to focuss 
our attention on possible
forms of $K(S+\bar S)$, exploring the chances of getting a 
global minimum at $V=0$. From eq.(\ref{condition}) we can check that 
a successful generalized dilaton dominated scenario (i.e. with $V=0$ at
${\rm Re}S\simeq 2$) requires very sizeable 
non-perturbative corrections to $K$. This can be seen by considering the 
form of eq.(\ref{condition}) when $K$ is substituted by the tree level 
expression $K=-\log(S+ \bar S)$. 
The fact that $a\simeq 18$ implies that the non-perturbative ``corrections''
to $K_{S},K_{\bar S S}$ must indeed be bigger than the tree level values.
If we further demand that the potential has a minimum at $V=0$ this implies
\bea
\label{Conds}
\left(-a+\frac{1}{2}K'\right)K''-\frac{3}{4}K'''&=& 0
\nonumber\\
\left(-a+\frac{1}{2}K'\right)^2-\frac{3}{4}K''&=& 0
\eea
(at the realistic point ${\rm Re}S=2$), where the primes denote derivatives
with respect to ${\rm Re}S$ (the second equation is simply 
eq.(\ref{condition})). It would be nice if the previous conditions
could be fulfilled  by some simple form of $K$.

\vspace{0.3cm}
\noindent
The fact that the SUGRA lagrangian has an exponential
dependence on $K$ (e.g. it can be written as 
${\cal L}=\left[e^{-K/3}\right]_D$) together with the 
fact that in the known examples (mainly orbifold compactifications) the 
perturbative corrections to $K$ are small, suggest that a sensible
decomposition for $K$ is 
\be
\label{K4}
e^{K}=e^{K_{tree}}+e^{K_{np}},
\ee
where the first term corresponds to the tree level expression 
($K_{tree}=-\log(2\ {\rm Re}S)$) and $K_{np}$ 
denotes the non-perturbative contributions.
The next step is to choose some plausible form for $K_{np}({\rm Re}S)$ and to
study the form of $V$, see eq.(\ref{V2}). 
The field theory contributions to 
$K_{np}$ have been evaluated in ref.\cite{mariano}, but for a realistic
case they turn out to be too tiny to appreciably modify the
tree-level results\footnote{ 
More precisely, the authors of ref.\cite{mariano} obtain
$K=-3\log\left[(2\ {\rm Re}S)^{1/3}+e^{-K_p/3}+
de^{-(2\ {\rm Re}S)/2c}\right]$, where $K_p$ denotes the perturbative
corrections to the tree-level expression, $d$ is an unknown constant
and $c$ is essentially the coefficient of the one-loop beta function
(the non-perturbative superpotential goes as $e^{-3S/2c}$).
}. On the other hand, 
according to the work of ref.\cite{Shenker}, stringy
non-perturbative effects may be sizeable (even for weak four-dimensional
gauge coupling) and 
plausibly go as $g^{-p}e^{-b/g}$ with
$p,b\sim O(1)$. Then a simple possibility is to take
\be
\label{K3}
e^{K_{np}}=d\ g^{-p}e^{-b/g}
\ee
where $d,p,b$ are constants with $p,b>0$ and $g^{-2}={\rm Re}S$. 

The numerical results indicate that, plugging this ansatz,
it is not difficult to get a minimum of the potential
using just one condensate with $a\simeq 18$ (thus guaranteeing the
correct size of SUSY breaking) and sensible values for the $d,p,b$ constants.
More precisely, for $d=-3$, $p=0$, $b=1$, there is a minimum of the potential
at ${\rm Re}S=2.1$. Unfortunately, the negative value of $d$ makes
this example unacceptable. Another example with positive $d$ is
$d=7.8$, $p=1$, $b=1$, which has a minimum at ${\rm Re}S=1.8$. However,
as a general result, playing just with these simple forms for $K$,
it seems impossible to get the minimum at $V=0$. Anyway, it is
impressive that just with one condensate and choosing very reasonable
values for $d$, $p$ and $b$ (note also that there is no fine-tuning in 
the previous choices), a minimum appears at the right value of the 
dilaton. Another typical characteristic of these examples is the appearance
of singularities in the potential caused by zeroes of the second derivative
of the K\"ahler potential, $K''$, at some particular values of $S$. 
Of course, this can be cured by additional terms in
$K$.

Finally, it would be worth to analyze the possible implications of 
this kind of non-perturbative K\"ahler potentials for other different matters,
such as the cosmological moduli problem \cite{moduli1,moduli2}.

\def\MPL #1 #2 #3 {{\em Mod.~Phys.~Lett.}~{\bf#1}\ (#2) #3 }
\def\NPB #1 #2 #3 {{\em Nucl.~Phys.}~{\bf B#1}\ (#2) #3 }
\def\PLB #1 #2 #3 {{\em Phys.~Lett.}~{\bf B#1}\ (#2) #3 }
\def\PR  #1 #2 #3 {{\em Phys.~Rep.}~{\bf#1}\ (#2) #3 }
\def\PRD #1 #2 #3 {{\em Phys.~Rev.}~{\bf D#1}\ (#2) #3 }
\def\PRL #1 #2 #3 {{\em Phys.~Rev.~Lett.}~{\bf#1}\ (#2) #3 }
\def\PTP #1 #2 #3 {{\em Prog.~Theor.~Phys.}~{\bf#1}\ (#2) #3 }
\def\RMP #1 #2 #3 {{\em Rev.~Mod.~Phys.}~{\bf#1}\ (#2) #3 }
\def\ZPC #1 #2 #3 {{\em Z.~Phys.}~{\bf C#1}\ (#2) #3 }

\end{document}